# Examining University Students' Artificial Intelligence-Generated Content (AIGC) Verification Intention from a Protection Motivation Perspective

**First Author and Corresponding Author**
Yiran Du
University of Cambridge, Cambridge, UK
yd392@cam.ac.uk

**Abstract**
Artificial Intelligence-Generated Content (AIGC) is increasingly used by students to support learning tasks, yet its outputs may contain inaccuracies, fabricated references, bias, and unsupported claims. This study examined students' intention to verify AIGC from the perspective of Protection Motivation Theory. A cross-sectional survey was conducted with 432 students who had experience using AIGC for learning. Structural equation modelling (SEM) was used to test the hypothesised relationships among threat appraisal, coping appraisal, protection motivation, and AIGC verification intention, while fuzzy-set qualitative comparative analysis (fsQCA) was applied to identify configurational pathways leading to high verification intention. The SEM results showed that protection motivation positively predicted AIGC verification intention. Perceived severity, perceived vulnerability, response efficacy, and self-efficacy positively influenced protection motivation, whereas maladaptive rewards and response cost had negative effects. The fsQCA results further revealed three configurations leading to high verification intention, with protection motivation appearing as a core condition across all pathways. These findings suggest that students' willingness to verify AIGC depends on both risk recognition and perceived coping capacity. The study extends Protection Motivation Theory to the context of AIGC verification and provides implications for promoting critical, responsible, and academically appropriate use of generative AI in higher education.

## 1. Introduction

Artificial Intelligence-Generated Content (AIGC) is increasingly embedded in higher education, where it supports students in information retrieval, writing, translation, summarisation, brainstorming, and problem-solving (Yan & Qianjun, 2025; Wang & Zhang, 2025). As generative AI systems become more widely used in learning contexts, AIGC has shifted from being a peripheral technological tool to a common source of academic support (Wu et al., 2026; Rizun et al., 2026). However, the fluent and authoritative style of AI-generated outputs may lead students to overestimate their reliability, despite the presence of hallucinated information, fabricated references, outdated knowledge, bias, and unsupported claims (Cui & Zhang, 2025; Huang et al., 2025).

Although prior research has examined AIGC adoption, learning benefits, and students' perceptions of generative AI in education, less attention has been paid to students' intention to verify AI-generated content before using it in learning tasks (Bai & Yang, 2025; Zhang et al., 2024). This gap is important because verification is closely associated with critical thinking, information literacy, academic integrity, and responsible engagement with digital information (Sun & Zhou, 2024; Zhao & Zhang, 2025). In the AIGC context, students' verification intention may depend not only on their awareness of risk, but also on whether they believe verification is effective, manageable, and worth the required effort.

To address this gap, the present study examines students' AIGC verification intention from the perspective of Protection Motivation Theory (PMT). Specifically, it investigates how threat appraisal factors, including perceived severity, perceived vulnerability, and maladaptive rewards, and coping appraisal factors, including response efficacy, self-efficacy, and response cost, shape protection motivation and, subsequently, AIGC verification intention (Rogers, 1975; Maddux & Rogers, 1983; Floyd et al., 2000). By combining structural equation modelling (SEM) and fuzzy-set qualitative comparative analysis (fsQCA), this study aims to identify both the net effects and configurational pathways through which students develop strong intentions to verify AIGC in higher education contexts.

## 2. Literature Review

### 2.1 Artificial Intelligence-Generated Content

Artificial Intelligence-Generated Content (AIGC) has attracted increasing scholarly attention as generative artificial intelligence becomes widely integrated into educational settings (Wu et al., 2026). AIGC generally refers to content produced by generative AI systems, including text, images, code, audio, video, and multimodal outputs (Rizun et al., 2026). In education, prior studies have highlighted the potential of AIGC to support students' learning by assisting with information retrieval, writing, translation, summarisation, brainstorming, and problem-solving (Yan & Qianjun, 2025). These affordances suggest that AIGC can improve learning efficiency and provide more flexible forms of academic support (J. Wang & Zhang, 2025).

However, the literature also points to substantial risks associated with AIGC use. Because AI-generated outputs are often fluent, coherent, and apparently authoritative, students may overestimate their reliability (Huang et al., 2025). Existing research has noted problems such as hallucinated information, fabricated references, embedded bias, outdated knowledge, and unsupported claims (Cui & Zhang, 2025). Therefore, recent educational research has increasingly shifted from asking whether students use AIGC to examining how they evaluate and regulate its use (Bai & Yang, 2025). This shift indicates that AIGC should be understood not only as a learning tool, but also as an information source that requires critical assessment and verification (X. Zhang et al., 2024).

### 2.2 Verification Intention

Verification intention has been examined in prior research on information behaviour, misinformation, digital literacy, and online risk management (Zhao & Zhang, 2025). It refers to an individual's willingness or planned tendency to check the accuracy, credibility, and reliability of information before accepting or using it (Qu & Lu, 2025). In digital learning environments, verification is particularly important because students frequently encounter information from multiple sources with varying levels of credibility (Ni & Huang, 2025). The emergence of AIGC further intensifies this issue, as AI-generated content may appear credible while containing factual inaccuracies or misleading claims (Armeen et al., 2024). Previous studies suggest that verification can include behaviours such as comparing information across sources, checking original references, consulting experts or peers, and evaluating the authority of information providers. (M. Chan, 2024; Kligler-Vilenchik, 2021; Ngo et al., 2023)

In educational contexts, students' verification intention is closely related to critical thinking, information literacy, and academic integrity (Sun & Zhou, 2024). Although AIGC can reduce effort and increase productivity, relying on unverified content may result in the reproduction of inaccurate knowledge and inappropriate academic practices (Stone, 2024). Accordingly, verification intention is an important construct for understanding whether students are prepared to engage with AIGC critically rather than accepting AI-generated outputs passively (Guo et al., 2024).

### 2.3 Protection Motivation Theory

Protection Motivation Theory (PMT) has been widely used to explain individuals' protective intentions and behaviours in response to perceived risks (Kiran et al., 2025). The theory proposes that protection motivation is shaped by two cognitive appraisal processes: threat appraisal and coping appraisal (Rogers, 1975). Threat appraisal concerns how individuals evaluate the risk itself, commonly including perceived severity, perceived vulnerability, and maladaptive rewards; coping appraisal concerns how individuals evaluate the protective action, including response efficacy, self-efficacy, and response cost (Maddux & Rogers, 1983). Prior studies have applied PMT beyond its original health-related context to areas such as cybersecurity, online privacy, misinformation avoidance, and technology use, demonstrating its usefulness for explaining protective decision-making in digital environments (Balla & Hagger, 2024; Kiran et al., 2025; Li et al., 2023).

In the context of AIGC, PMT provides a relevant theoretical lens because students may judge both the potential consequences of relying on inaccurate AI-generated content and the feasibility of verifying such content. If students perceive unverified AIGC as risky and believe that verification is effective and manageable, they are more likely to form stronger verification intentions. Therefore, PMT offers a structured basis for reviewing and explaining the psychological mechanisms underlying students' intention to verify AIGC in educational contexts.

## 3. The Conceptual Model and Hypotheses

The conceptual model developed in this study (see Figure 1) is grounded in Protection Motivation Theory and explains students' AIGC verification intention through two appraisal processes: threat appraisal and coping appraisal. Threat appraisal includes perceived severity, perceived vulnerability, and maladaptive rewards, reflecting students' assessment of the risks and benefits associated with not verifying AIGC. Coping appraisal includes response efficacy, self-efficacy, and response cost, reflecting students' evaluation of the effectiveness, feasibility, and burden of verification. Together, these factors are expected to shape students' protection motivation, which in turn influences their intention to verify AIGC.

**Figure 1. The Conceptual Model**

Threat Appraisal
Perceived Severity
Perceived Vulnerability
Maladaptive Rewards
H2 (+)
H3 (+)
H4 (-)
H5 (+)
H6 (+)
H7 (-)
Protection Motivation
H1 (+)
AIGC Verification Intention
Response Efficacy
Self-Efficacy
Response Cost
Coping Appraisal

### 3.1 Protection Motivation

Protection motivation refers to an individual's intention to adopt protective actions when facing a perceived threat (Hinssen & Dohle, 2023). According to Protection Motivation Theory (PMT), protection motivation is formed through threat appraisal and coping appraisal, and it functions as the proximal psychological driver of protective behaviour (Floyd et al., 2000). In this study, protection motivation represents students' overall motivation to protect themselves from the potential risks of inaccurate, misleading, or unreliable AIGC. Theoretically, when students recognise that unverified AIGC may harm their learning quality, academic performance, or academic integrity, they are more likely to develop an intention to verify AI-generated content before using it (X. Zou et al., 2023). Empirically, PMT has been widely applied to explain protective intentions and behaviours in digital contexts, including information security, online privacy, and technology-related risk management, supporting the role of protection motivation as a central predictor of protective behavioural intention (Cheng, 2025; Karayel & Saygili, 2025; Kiran et al., 2025). Therefore, this study proposes the following hypothesis:

H1: Protection motivation positively affects students' AIGC verification intention.

### 3.2 Perceived Severity

Perceived severity in this study refers to students' assessment of the seriousness of the negative consequences associated with using unverified AIGC. In the AIGC-supported learning context, these consequences may include misunderstanding knowledge, relying on fabricated references, producing inaccurate assignments, or violating academic integrity requirements (Abbas et al., 2024). From a PMT perspective, individuals are more likely to engage in protective cognition when they perceive the potential consequences of a threat as serious (Maddux & Rogers, 1983). Empirical studies based on PMT have shown that perceived severity is positively associated with protection motivation and protective behavioural intention across different risk contexts, including health and information security behaviours (Ifinedo, 2012; Siponen et al., 2014; Sommestad et al., 2015). Therefore, when students perceive the consequences of unverified AIGC use as severe, they are more likely to be motivated to verify AI-generated content.

H2: Perceived severity positively affects students' protection motivation.

### 3.3 Perceived Vulnerability

Perceived vulnerability in this study refers to students' perceived likelihood of being affected by the risks of unverified AIGC. Students may feel vulnerable when they frequently use AIGC, lack sufficient domain knowledge, or find it difficult to distinguish accurate information from hallucinated or misleading AI-generated content (Stone, 2024). PMT suggests that individuals are more likely to form protection motivation when they believe that a threat is personally relevant and likely to affect them (Maddux & Rogers, 1983). Prior empirical research has found that perceived vulnerability is an important antecedent of protective intention, particularly in contexts where individuals must judge their exposure to uncertain digital or informational risks (Hanus & Wu, 2015; Ifinedo, 2012; Karayel & Saygili, 2025). In this study, students who perceive themselves as more vulnerable to AIGC-related risks are expected to show stronger motivation to verify such content.

H3: Perceived vulnerability positively affects students' protection motivation.

### 3.4 Maladaptive Rewards

Maladaptive rewards in this study refer to the perceived benefits of not adopting protective behaviour. In the context of AIGC use, students may avoid verification because accepting AI-generated content directly can save time, reduce cognitive effort, and improve task efficiency (Yan & Qianjun, 2025). Although these benefits are attractive, PMT argues that maladaptive rewards weaken protection motivation because they make risky behaviour appear more convenient or beneficial (Rogers, 1975). In digital contexts, empirical studies have similarly shown that users may avoid protective actions when unsafe or less cautious behaviours provide immediate convenience or efficiency benefits (Basarmak & Ates, 2026; Cheng, 2025; Kiran et al., 2025). Therefore, students who perceive stronger rewards from not verifying AIGC are less likely to develop motivation to engage in verification.

H4: Maladaptive rewards negatively affect students' protection motivation.

### 3.5 Response Efficacy

Response efficacy in this study refers to students' belief that verifying AIGC is effective in reducing the risks associated with inaccurate or unreliable AI-generated content. Verification may include checking original sources, comparing information across platforms, consulting teachers or peers, and evaluating the credibility of cited references (Guo et al., 2024). PMT argues that individuals are more likely to adopt protective actions when they believe the recommended response can effectively reduce the threat (Floyd et al., 2000). Empirical PMT research has consistently identified response efficacy as a positive predictor of protection motivation and protective intention in health, information security, and technology-related risk contexts (Balla & Hagger, 2024; Janmaimool, 2017; Siponen et al., 2014).

Accordingly, when students believe that verification can help them identify errors and improve the reliability of AIGC use, they are more likely to be motivated to verify AI-generated content.

H5: Response efficacy positively affects students' protection motivation.

### 3.6 Self-Efficacy

Self-efficacy refers in this study to students' confidence in their ability to verify AIGC effectively. Students with high self-efficacy are more likely to believe that they can identify unreliable information, evaluate source credibility, compare competing claims, and use appropriate verification strategies (C. K. Y. Chan & Hu, 2023). Within PMT, self-efficacy is a key coping appraisal construct because individuals are unlikely to adopt protective actions if they do not believe they can perform them successfully (Rogers, 1975). Empirical studies have repeatedly shown that self-efficacy is one of the strongest predictors of protective motivation and behavioural intention, particularly in contexts requiring individual judgement and active response (Karayel & Saygili, 2025; Nabizadeh et al., 2018; Tsai et al., 2016). Therefore, students with stronger confidence in their verification ability are expected to show higher protection motivation.

H6: Self-efficacy positively affects students' protection motivation.

### 3.7 Response Cost

Response cost refers to the perceived time, effort, and resources required to verify AIGC. Although verification can reduce the risks of using inaccurate AI-generated content, students may perceive it as burdensome because it requires additional reading, source comparison, reference checking, or consultation with others (Sun & Zhou, 2024). PMT proposes that higher response cost weakens protection motivation because individuals are less likely to adopt protective actions when they view them as difficult, time-consuming, or inconvenient (Maddux & Rogers, 1983). Empirical research in information security and technology use has similarly found that perceived cost can reduce users' willingness to engage in protective or responsible behaviours (Eitze et al., 2025; Hinssen & Dohle, 2023; Nabizadeh et al., 2018). Therefore, students who perceive verification as costly are expected to have lower protection motivation.

H7: Response cost negatively affects students' protection motivation.

## 4. Methodology

### 4.1 Research Design

This study adopted a cross-sectional survey design to examine students' intention to verify artificial intelligence-generated content (AIGC) from the perspective of Protection Motivation Theory (PMT). Self-reported questionnaire data were collected from students who had experience using AIGC in learning contexts. The research model specified perceived severity, perceived vulnerability, maladaptive rewards, response efficacy, self-efficacy, and response cost as antecedents of protection motivation, and protection motivation as the proximal predictor of AIGC verification intention. This design was appropriate because the study aimed to test theoretically derived relationships among latent psychological constructs. Structural equation modelling (SEM) was used to assess the measurement model and test the hypothesised net effects, while fuzzy-set qualitative comparative analysis (fsQCA) was subsequently employed to identify configurational pathways leading to high AIGC verification intention.

### 4.2 Participants

Participants were recruited through Credamo, a professional online survey platform widely used for questionnaire-based academic research in China. To be eligible for the study, participants were required to be students and to have prior experience using artificial intelligence-generated content (AIGC) for learning purposes. Before data analysis, the collected responses were screened to ensure data quality. Responses were excluded if participants did not meet the eligibility criteria, submitted incomplete questionnaires, failed attention-check items, or showed evidence of careless responding, such as patterned answers or unusually short completion times (Ward & Meade, 2023). After this exclusion process, 432 valid responses were retained for analysis. As shown in Table 1, the final sample included 194 male students (44.9%) and 238 female students (55.1%). In terms of age, 156 participants were aged 18–20 years (36.1%), 201 were aged 21–23 years (46.5%), and 75 were aged 24 or above (17.4%). Regarding study level, 287 participants were undergraduates (66.4%) and 145 were postgraduates (33.6%). The sample was also relatively balanced by academic discipline, with 213 students from STEM disciplines (49.3%) and 219 from non-STEM disciplines (50.7%). These characteristics indicate that the retained sample covered students with diverse demographic and academic backgrounds, providing an appropriate basis for examining AIGC verification intention in higher education contexts.

**Table 1. Participant Characteristics (*N* = 432)**

| Characteristic | Category | *n* | % |
|---|---|---|---|
| Gender | Male | 194 | 44.9 |
| | Female | 238 | 55.1 |
| Age | 18–20 | 156 | 36.1 |
| | 21–23 | 201 | 46.5 |
| | 24 or above | 75 | 17.4 |
| Study level | Undergraduate | 287 | 66.4 |
| | Postgraduate | 145 | 33.6 |
| Academic discipline | STEM | 213 | 49.3 |
| | Non-STEM | 219 | 50.7 |

### 4.3 Measurement

All constructs were measured using multi-item scales adapted from prior research on Protection Motivation Theory and information verification behaviour (Balla & Hagger, 2024; Eitze et al., 2025; Hinssen & Dohle, 2023; Karayel & Saygili, 2025; Kiran et al., 2025; Li et al., 2023), with wording revised to fit the context of artificial intelligence-generated content (AIGC) use in learning (Y. Du, 2024). The questionnaire measured eight latent constructs: AIGC verification intention, protection motivation, perceived severity, perceived vulnerability, maladaptive rewards, response efficacy, self-efficacy, and response cost. Each construct was assessed with three items, resulting in 24 measurement items in total, as shown in Table 2. Responses were collected using a five-point Likert scale ranging from 1 = strongly disagree to 5 = strongly agree, with higher scores indicating stronger agreement with the corresponding construct. The original English items were translated into Chinese and then back-translated into English by bilingual researchers to ensure linguistic accuracy and conceptual equivalence (Klotz et al., 2023). Discrepancies were discussed and resolved before the formal survey. A pilot test was also conducted with a small group of students (*N* = 30) to assess item clarity, wording, and questionnaire flow, and minor revisions were made based on their feedback. Because all data were collected through a self-report questionnaire, common method variance (CMV) was assessed (Podsakoff et al., 2024). Harman's single-factor test showed that the first unrotated factor accounted for less than 50% of the total variance, suggesting that CMV was not a serious concern (Aguirre-Urreta & Hu, 2019). The reliability and validity of the measurement model were subsequently assessed through confirmatory factor analysis, internal consistency reliability, convergent validity, and discriminant validity.

**Table 2. Constructs and Measurement Items**

| Construct | Item | Measurement Item (English) | Measurement Item (Chinese) |
|---|---|---|---|
| AIGC Verification Intention | VI1 | I intend to verify the accuracy of AIGC before using it for learning tasks. | 我打算在将 AIGC 用于学习任务前核实其准确性。 |
| | VI2 | I am willing to check whether AIGC is reliable before accepting its content. | 我愿意在接受 AIGC 内容前检查其可靠性。 |
| | VI3 | I plan to compare AIGC with other reliable sources before using it. | 我计划在使用 AIGC 前将其与其他可靠来源进行比较。 |
| Protection Motivation | PM1 | I feel motivated to protect myself from the risks of using unverified AIGC. | 我有动力保护自己免受使用未经核实的 AIGC 所带来的风险。 |
| | PM2 | I think it is necessary to take action to avoid the negative effects of inaccurate AIGC. | 我认为有必要采取行动以避免不准确 AIGC 带来的负面影响。 |
| | PM3 | I am motivated to use AIGC cautiously in my learning activities. | 我有动力在学习活动中谨慎使用 AIGC。 |
| Perceived Severity | PS1 | Using inaccurate AIGC may seriously affect the quality of my learning outcomes. | 使用不准确的 AIGC 可能会严重影响我的学习成果质量。 |
| | PS2 | Relying on unverified AIGC may lead to serious academic problems. | 依赖未经核实的 AIGC 可能会导致严重的学术问题。 |
| | PS3 | The consequences of using misleading AIGC in academic work can be serious. | 在学术任务中使用误导性的 AIGC 可能会产生严重后果。 |
| Perceived Vulnerability | PV1 | I may encounter inaccurate or misleading information when using AIGC. | 我在使用 AIGC 时可能会遇到不准确或误导性的信息。 |
| | PV2 | I am likely to be affected by the risks of unverified AIGC. | 我可能会受到未经核实 AIGC 所带来的风险影响。 |
| | PV3 | It is possible that I may mistakenly trust incorrect AIGC. | 我有可能错误地信任不正确的 AIGC 内容。 |
| Maladaptive Rewards | MR1 | Not verifying AIGC can save me time. | 不核实 AIGC 可以节省我的时间。 |
| | MR2 | Directly using AIGC without verification can reduce my effort. | 不经核实直接使用 AIGC 可以减少我的精力投入。 |
| | MR3 | Accepting AIGC without checking makes learning tasks more convenient. | 不经检查就接受 AIGC 会使学习任务更加方便。 |
| Response Efficacy | RE1 | Verifying AIGC can help me identify inaccurate information. | 核实 AIGC 可以帮助我识别不准确的信息。 |
| | RE2 | Checking AIGC against reliable sources can reduce the risks of using incorrect content. | 将 AIGC 与可靠来源进行核对可以降低使用错误内容的风险。 |
| | RE3 | Verification is an effective way to improve the reliability of AIGC use. | 核实是提高 AIGC 使用可靠性的有效方式。 |
| Self-Efficacy | SE1 | I am confident that I can verify the accuracy of AIGC. | 我有信心核实 AIGC 的准确性。 |
| | SE2 | I have the ability to judge whether AIGC is reliable. | 我有能力判断 AIGC 是否可靠。 |
| | SE3 | I can use appropriate strategies to check AIGC when necessary. | 必要时，我能够使用适当的方法检查 AIGC。 |
| Response Cost | RC1 | Verifying AIGC requires too much time. | 核实 AIGC 需要花费太多时间。 |
| | RC2 | Checking the reliability of AIGC requires considerable effort. | 检查 AIGC 的可靠性需要投入较多精力。 |
| | RC3 | It is inconvenient for me to verify AIGC before using it. | 在使用 AIGC 前进行核实对我来说不太方便。 |

### 4.4 Data Analysis

Data analysis was conducted in two stages. First, structural equation modelling (SEM) (Kline, 2023) was used to assess the measurement model and test the hypothesised relationships among the PMT constructs. Descriptive statistics, normality, model fit, reliability, convergent validity, and discriminant validity were examined before testing the structural paths and the mediating role of protection motivation. Second, fuzzy-set qualitative comparative analysis (fsQCA) (Schneider & Wagemann, 2012) was employed to identify different configurations of threat appraisal, coping appraisal, and protection motivation conditions that could lead to high AIGC verification intention. The constructs were calibrated into fuzzy sets using the 95th, 50th, and 5th percentiles as anchors for full membership, crossover, and full non-membership, respectively. Necessity and sufficiency analyses were then performed to examine whether individual conditions or combinations of conditions explained students' high verification intention.

## 5. Results

### 5.1 SEM Results

The descriptive statistics, model fit, reliability, convergent validity, and discriminant validity results are presented in Tables 3–7. As shown in Table 3, the mean scores of the constructs ranged from 3.21 to 3.82, suggesting that students generally reported moderate to relatively high perceptions across the PMT-related variables. Skewness and kurtosis values were within acceptable ranges, indicating no serious deviation from normality. The model fit indices in Table 4 showed that both the measurement model and the structural model achieved acceptable fit, with $\chi^2$/df values below 3.00, CFI and TLI values above 0.90, and RMSEA and SRMR values below 0.08. Reliability and convergent validity were also satisfactory, as shown in Table 5. All factor loadings exceeded 0.70, Cronbach's $\alpha$ values ranged from 0.81 to 0.85, composite reliability values ranged from 0.82 to 0.86, and AVE values ranged from 0.60 to 0.67. These results indicate that the measurement items had adequate internal consistency and convergent validity. In addition, the Fornell–Larcker results in Table 6 and the HTMT ratios in Table 7 supported discriminant validity, as the square roots of AVE were higher than the inter-construct correlations and all HTMT values were below the recommended threshold.

**Table 3. Descriptive Statistics of the Constructs**

| Construct | *M* | *SD* | Skewness | Kurtosis |
|---|---|---|---|---|
| AIGC Verification Intention | 3.21 | 0.86 | -0.12 | -0.41 |
| Protection Motivation | 3.34 | 0.82 | -0.18 | -0.29 |
| Perceived Severity | 3.56 | 0.79 | -0.24 | -0.18 |
| Perceived Vulnerability | 3.48 | 0.84 | -0.19 | -0.26 |
| Maladaptive Rewards | 3.82 | 0.76 | -0.36 | 0.08 |
| Response Efficacy | 3.71 | 0.77 | -0.31 | -0.05 |
| Self-Efficacy | 3.45 | 0.81 | -0.17 | -0.22 |
| Response Cost | 3.69 | 0.78 | -0.28 | -0.09 |

**Table 4. Model Fit Indices**

| Fit Index | Threshold | Measurement Model | Structural Model |
|---|---|---|---|
| $\chi^2$/df | < 3.00 | 2.61 | 2.78 |
| CFI | > 0.90 | 0.94 | 0.93 |
| TLI | > 0.90 | 0.93 | 0.92 |
| RMSEA | < 0.08 | 0.06 | 0.07 |
| SRMR | < 0.08 | 0.05 | 0.06 |

**Table 5. Reliability and Convergent Validity**

| Construct | Item | Loading | Cronbach's α | CR | AVE |
|---|---|---|---|---|---|
| AIGC Verification Intention | VI1 | 0.78 | 0.84 | 0.85 | 0.66 |
| | VI2 | 0.83 | | | |
| | VI3 | 0.82 | | | |
| Protection Motivation | PM1 | 0.77 | 0.83 | 0.84 | 0.64 |
| | PM2 | 0.81 | | | |
| | PM3 | 0.82 | | | |
| Perceived Severity | PS1 | 0.75 | 0.82 | 0.83 | 0.62 |
| | PS2 | 0.80 | | | |
| | PS3 | 0.81 | | | |
| Perceived Vulnerability | PV1 | 0.76 | 0.83 | 0.84 | 0.63 |
| | PV2 | 0.81 | | | |
| | PV3 | 0.80 | | | |
| Maladaptive Rewards | MR1 | 0.74 | 0.81 | 0.82 | 0.60 |
| | MR2 | 0.79 | | | |
| | MR3 | 0.79 | | | |
| Response Efficacy | RE1 | 0.79 | 0.85 | 0.86 | 0.67 |
| | RE2 | 0.83 | | | |
| | RE3 | 0.83 | | | |
| Self-Efficacy | SE1 | 0.76 | 0.83 | 0.84 | 0.64 |
| | SE2 | 0.81 | | | |
| | SE3 | 0.82 | | | |
| Response Cost | RC1 | 0.75 | 0.82 | 0.83 | 0.62 |
| | RC2 | 0.80 | | | |
| | RC3 | 0.81 | | | |

**Table 6. Discriminant Validity (Fornell–Larcker Criterion)**

| Construct | VI | PM | PS | PV | MR | RE | SE | RC |
|---|---|---|---|---|---|---|---|---|
| VI | **0.81** | | | | | | | |
| PM | 0.54 | **0.80** | | | | | | |
| PS | 0.39 | 0.48 | **0.79** | | | | | |
| PV | 0.37 | 0.46 | 0.42 | **0.79** | | | | |
| MR | -0.28 | -0.35 | -0.24 | -0.27 | **0.77** | | | |
| RE | 0.45 | 0.52 | 0.36 | 0.38 | -0.30 | **0.82** | | |
| SE | 0.43 | 0.50 | 0.34 | 0.36 | -0.26 | 0.44 | **0.80** | |
| RC | -0.31 | -0.39 | -0.29 | -0.32 | 0.41 | -0.33 | -0.35 | **0.79** |

Note. Diagonal elements in bold represent the square root of the average variance extracted (AVE).

**Table 7. Discriminant Validity (HTMT Ratio)**

| Construct | VI | PM | PS | PV | MR | RE | SE | RC |
|---|---|---|---|---|---|---|---|---|
| VI | — | | | | | | | |
| PM | 0.64 | — | | | | | | |
| PS | 0.47 | 0.57 | — | | | | | |
| PV | 0.44 | 0.55 | 0.50 | — | | | | |
| MR | 0.34 | 0.42 | 0.30 | 0.33 | — | | | |
| RE | 0.53 | 0.62 | 0.43 | 0.46 | 0.36 | — | | |
| SE | 0.51 | 0.60 | 0.41 | 0.44 | 0.32 | 0.52 | — | |
| RC | 0.37 | 0.47 | 0.35 | 0.39 | 0.49 | 0.40 | 0.42 | — |

The structural model results are reported in Table 8 and Figure 2. Protection motivation had a significant positive effect on AIGC verification intention ($\beta = 0.52$, $z = 8.67$), supporting H1. For the antecedents of protection motivation, perceived severity ($\beta = 0.19$, $z = 3.17$), perceived vulnerability ($\beta = 0.17$, $z = 2.83$), response efficacy ($\beta = 0.24$, $z = 4.00$), and self-efficacy ($\beta = 0.22$, $z = 3.67$) exerted significant positive effects, supporting H2, H3, H5, and H6. In contrast, maladaptive rewards ($\beta = -0.21$, $z = -4.20$) and response cost ($\beta = -0.18$, $z = -3.00$) had significant negative effects on protection motivation, supporting H4 and H7. The mediation analysis in Table 9 further showed that protection motivation significantly mediated the relationships between all six antecedent variables and AIGC verification intention, as the bootstrapped 95% confidence intervals did not include zero. Specifically, positive indirect effects were found for perceived severity, perceived vulnerability, response efficacy, and self-efficacy, whereas negative indirect effects were found for maladaptive rewards and response cost.

Overall, the SEM results provide empirical support for the proposed PMT-based model and indicate that students' AIGC verification intention is shaped by both threat appraisal and coping appraisal through protection motivation.

**Table 8. Structural Model Results**

| Hypothesis | Path | $\beta$ | SE | $z$ | Result |
|---|---|---|---|---|---|
| H1 | PM → VI | 0.52*** | 0.06 | 8.67 | Supported |
| H2 | PS → PM | 0.19** | 0.06 | 3.17 | Supported |
| H3 | PV → PM | 0.17** | 0.06 | 2.83 | Supported |
| H4 | MR → PM | -0.21*** | 0.05 | -4.20 | Supported |
| H5 | RE → PM | 0.24*** | 0.06 | 4.00 | Supported |
| H6 | SE → PM | 0.22*** | 0.06 | 3.67 | Supported |
| H7 | RC → PM | -0.18** | 0.06 | -3.00 | Supported |

Note. *** $p < 0.001$; ** $p < 0.01$.

**Figure 2. Structural Model Results**

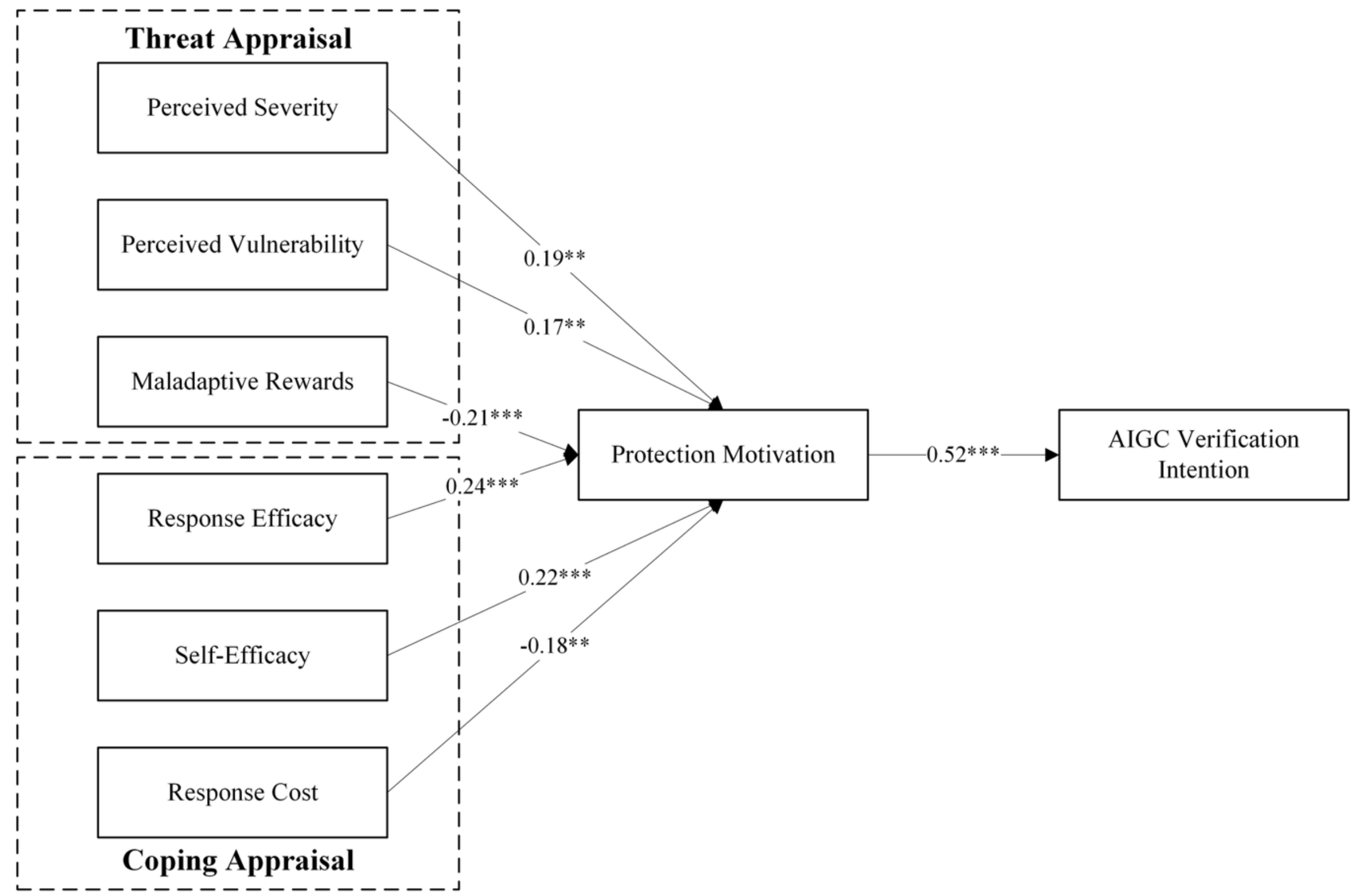

Note. *** $p < 0.001$; ** $p < 0.01$.

**Table 9. Mediation Analysis Results (Bootstrapping)**

| Indirect Path | $\beta$ | SE | 95% CI | Result |
|---|---|---|---|---|
| PS → PM → VI | 0.10 | 0.03 | [0.04, 0.17] | Supported |
| PV → PM → VI | 0.09 | 0.03 | [0.03, 0.15] | Supported |
| MR → PM → VI | -0.11 | 0.03 | [-0.18, -0.06] | Supported |
| RE → PM → VI | 0.12 | 0.03 | [0.06, 0.20] | Supported |
| SE → PM → VI | 0.11 | 0.03 | [0.05, 0.18] | Supported |
| RC → PM → VI | -0.09 | 0.03 | [-0.16, -0.03] | Supported |

### 5.2 fsQCA Results

The fsQCA used perceived severity, perceived vulnerability, maladaptive rewards, response efficacy, self-efficacy, response cost, and protection motivation as condition variables, with AIGC verification intention as the outcome. These variables were selected based on the PMT-based conceptual model and the preceding SEM analysis. Before analysis, all constructs were calibrated into fuzzy-set scores using three anchors: the 95th percentile for full membership, the 50th percentile for the crossover point, and the 5th percentile for full non-membership. A necessity analysis was then conducted. No single condition reached the conventional consistency threshold of 0.90, indicating that high AIGC verification intention was not dependent on any individual antecedent condition alone.

The sufficiency analysis identified three configurations leading to high AIGC verification intention (see Table 10). The consistency values for the three paths were 0.89, 0.86, and 0.84, with an overall consistency of 0.87 and overall coverage of 0.62. Protection motivation appeared as a core condition across all three paths, suggesting its central role in students' verification intention. Path 1 reflects a "high threat–high coping–low cost" configuration, combining perceived severity, response efficacy, protection motivation, low maladaptive rewards, and low response cost. Path 2 represents an "efficacy-driven protection" configuration, centred on response efficacy and protection motivation. Path 3 reflects a "vulnerability–self-efficacy protection" configuration, combining perceived vulnerability, self-efficacy, protection motivation, response efficacy, and low response cost. Overall, the results indicate equifinality: students' high AIGC verification intention can arise from different combinations of threat appraisal, coping appraisal, and protection motivation conditions.

**Table 10. Configuration Paths of AIGC Verification Intention**

| **Condition** | **Path 1** | **Path 2** | **Path 3** |
|---|---|---|---|
| Perceived severity | ● | ○ | |
| Perceived vulnerability | ○ | | ● |
| Maladaptive rewards | ⊖ | ⊗ | |
| Response efficacy | ● | ● | ○ |
| Self-efficacy | ○ | | ● |
| Response cost | ⊗ | | ⊖ |
| Protection motivation | ● | ● | ● |
| Consistency | 0.89 | 0.86 | 0.84 |
| Raw coverage | 0.34 | 0.28 | 0.25 |
| Unique coverage | 0.12 | 0.09 | 0.07 |
| Overall consistency | 0.87 | | |
| Overall coverage | 0.62 | | |

Note. ● indicates the presence of a core condition; ○ indicates the presence of a peripheral condition; ⊗ indicates the absence of a core condition; ⊖ indicates the absence of a peripheral condition; blank cells indicate "don't care" conditions.

## 6. Discussion

### 6.1 Net Effects on AIGC Verification Intention

The SEM results indicate that protection motivation is a significant direct predictor of students' AIGC verification intention. This finding supports the core assumption of Protection Motivation Theory that protective behavioural intention is shaped by individuals' motivation to avoid or reduce perceived risks (Rogers, 1975; Maddux & Rogers, 1983; Floyd et al., 2000). In this study, students who were more motivated to protect themselves from the risks of inaccurate, misleading, or unreliable AIGC were more likely to intend to verify AIGC before using it for learning tasks. This result is consistent with prior PMT-based research in digital and technology-related contexts, which has shown that protection motivation plays a central role in explaining protective behavioural intentions (Kiran et al., 2025; Karayel & Saygili, 2025; Cheng, 2025). It also extends existing work on AIGC and information verification by showing that students' verification intention is not merely a matter of general digital literacy, but is closely related to their perceived need for self-protection in AI-supported learning environments (Guo et al., 2024; Zhao & Zhang, 2025).

The antecedent effects further show that both threat appraisal and coping appraisal contribute to protection motivation. Perceived severity and perceived vulnerability had significant positive effects, suggesting that students are more motivated to verify AIGC when they recognise the seriousness of potential academic consequences and perceive themselves as susceptible to inaccurate or misleading AI-generated content. This aligns with PMT and prior studies showing that risk perception can strengthen protective motivation (Ifinedo, 2012; Hanus & Wu, 2015; Sommestad et al., 2015). Response efficacy and self-efficacy also positively predicted protection motivation, indicating that students are more likely to form verification intentions when they believe that verification is effective and that they are capable of carrying it out (Balla & Hagger, 2024; Janmaimool, 2017; Tsai et al., 2016). By contrast, maladaptive rewards and response cost negatively affected protection motivation, implying that the convenience of not verifying AIGC and the perceived burden of verification may weaken students' willingness to engage in protective behaviour. This finding is consistent with previous research showing that immediate convenience and perceived cost can discourage protective action in digital contexts (Hinssen & Dohle, 2023; Eitze et al., 2025; Kiran et al., 2025).

### 6.2 Configurational Mechanisms of AIGC Verification Intention

The fsQCA results complement the SEM findings by showing that students' high AIGC verification intention can be produced through multiple configurational pathways rather than a single linear mechanism. No individual condition was necessary for high verification intention, indicating that students do not rely on one isolated factor when deciding whether to verify AIGC. Instead, high verification intention emerges from combinations of threat appraisal, coping appraisal, and protection motivation. This supports the principle of equifinality in configurational analysis, where different combinations of conditions may lead to the same outcome (Schneider & Wagemann, 2012). Notably, protection motivation appeared as a core condition in all three configurations, confirming its central role in linking students' risk perceptions and coping beliefs to verification intention. This result is consistent with Protection Motivation Theory, which positions protection motivation as the proximal driver of protective behavioural intention (Rogers, 1975; Floyd et al., 2000), and with prior studies that identify protection motivation as a key mechanism in digital risk-related behaviours (Kiran et al., 2025; Karayel & Saygili, 2025).

The three configurations further reveal distinct mechanisms underlying high AIGC verification intention. Path 1, characterised by high perceived severity, high response efficacy, strong protection motivation, low maladaptive rewards, and low response cost, suggests that students are most likely to verify AIGC when they perceive serious consequences, believe verification is effective, and do not see non-verification as convenient or verification as burdensome. Path 2 reflects an efficacy-driven mechanism in which response efficacy and protection motivation are sufficient core conditions, indicating that students may form strong verification intention even when threat perceptions are not dominant, provided that they believe verification can effectively reduce the risks of unreliable AIGC. Path 3 represents a vulnerability–self-efficacy mechanism, where perceived vulnerability, self-efficacy, response efficacy, protection motivation, and low response cost jointly support high verification intention. This suggests that students who feel exposed to AIGC-related risks are more willing to verify content when they also feel capable of doing so and perceive the verification process as manageable. Together, these pathways show that AIGC verification intention is shaped by different but functionally equivalent patterns of risk recognition, efficacy belief, and cost evaluation, which is consistent with prior research on information verification and protective behaviour in technology-mediated environments (Chan, 2024; Ngo et al., 2023; Ni & Huang, 2025).

### 6.3 Theoretical and Practical Implications

This study extends Protection Motivation Theory to AIGC verification in higher education by showing that students' verification intention is shaped by both threat appraisal and coping appraisal. The SEM results explain the net effects of individual factors, while the fsQCA results reveal multiple configurational pathways to high verification intention. Together, these findings suggest that students' AIGC verification intention is not driven by a single factor, but by different combinations of perceived risk, efficacy beliefs, perceived cost, and protection motivation. This broadens research on AIGC use by shifting attention from adoption and learning efficiency to critical and responsible engagement with AI-generated content.

Practically, the findings suggest that educators should help students recognise the risks of unverified AIGC while also providing feasible verification strategies. Risk awareness alone may be insufficient; students also need confidence and practical skills to check AI-generated content effectively. Higher education institutions could integrate AIGC verification training into digital literacy, academic writing, and research methods courses. To reduce response cost, they should provide simple tools such as verification checklists, source-evaluation guides, examples of fabricated references, and discipline-specific verification procedures.

**6.4 Limitations and Future Directions**

This study has several limitations. First, it used a cross-sectional survey design, which limits the ability to infer causal relationships among the variables (Y. Du, Tang, et al., 2026; Y. Du, Yuan, et al., 2026; Y. Du & He, 2026e, 2026c; Jia et al., 2026; Tang, Jia, et al., 2026; Tang, Lau, et al., 2026; C. Wang, Du, et al., 2026; C. Wang, Zou, et al., 2026; W. Zhang et al., 2026). Future studies could adopt longitudinal or experimental designs to examine how students' AIGC verification intention changes over time or in response to specific instructional interventions (Chen et al., 2022; C. Du et al., 2025; Y. Du, 2023, 2024, 2025b, 2025a, 2026; Y. Du et al., 2024, 2025; Y. Du, Li, et al., 2026; Y. Du & He, 2026d, 2026b, 2026a; He & Du, 2024; C. Wang et al., 2024; B. Zou et al., 2023, 2024). Second, the data were based on self-reported measures, which may be affected by social desirability or response bias. Future research could include behavioural measures, such as students' actual verification actions when using AIGC for learning tasks.

Third, although the sample included students from different study levels and academic disciplines, it was collected in one national context. Future studies could examine whether the proposed model applies across different cultural, institutional, and educational settings. Finally, this study focused on general AIGC verification intention rather than specific verification practices. Future research could distinguish between different verification behaviours, such as checking references, comparing sources, using fact-checking tools, or consulting teachers and peers.

**7. Conclusion**

This study examined students' AIGC verification intention from the perspective of Protection Motivation Theory. The SEM results showed that protection motivation directly promoted verification intention, while perceived severity, perceived vulnerability, response efficacy, and self-efficacy strengthened protection motivation, and maladaptive rewards and response cost weakened it. The fsQCA results further revealed that high verification intention can emerge through multiple configurations of threat appraisal, coping appraisal, and protection motivation. Overall, the findings suggest that students' willingness to verify AIGC depends not only on recognising its potential risks, but also on believing that verification is effective, manageable, and worth the effort.